# Leveraging Plasmonic Nanocavity Arrays Forming Metasurfaces to Boost Second Harmonic Generation due to Surface Effects

Sky Semone,[1,2] Matthew J. Brandsema,[2] Thang Hoang[3], and Christos Argyropoulos[1,*]

[1]Department of Electrical Engineering, Pennsylvania State University, University Park, PA, 16803 USA
[2]Applied Research Labs, Pennsylvania State University, University Park, PA, 16803 USA
[3]Department of Physics and Material Science, University of Memphis, Memphis, TN, 38152 USA
*cfa5361@psu.edu

## Abstract

Plasmonic metasurfaces have emerged as a promising platform for enhancing a range of nonlinear optical processes, offering compact geometry and flexibility in light manipulation. Second order nonlinear processes, like second harmonic generation (SHG), typically require non-centrosymmetric crystals to be realized. Here, we experimentally demonstrate enhanced SHG response by using a gold nanocavity array forming a plasmonic metasurface absorber where titanium dioxide ($TiO_2$), a centrosymmetric dielectric material, with subwavelength thickness is deposited in the realized nanogaps. While such dielectric material has an extremely low second order nonlinear susceptibility, we observe $10^5$-fold boosting in the nonlinear SHG process mainly due to the surface nonlinear susceptibility of the gold metal aided by the significant electric field enhancement that occurs in the nanogaps due to the formed nanocavity resonance. The experimental results obtained are theoretically explained with extensive and rigorous nonlinear simulations that consider all the bulk and surface linear and nonlinear material properties. The presented robust harmonic generation from an ultrathin plasmonic metasurface can be used in nonlinear and quantum integrated photonic applications.

# 1. Introduction

Second harmonic generation (SHG), a second-order nonlinear optical process that converts incident photons at fundamental frequency $\omega$ into photons at second harmonic frequency $2\omega$, has established itself as a key optical mechanism in modern photonics.[1,2] Motivated by the demand for highly integrated photonic circuitry, the past few decades have seen substantial progress in miniaturizing SHG platforms to the nanoscale, facilitating the realization of ultracompact and efficient nonlinear light sources. Increasing the efficiency of SHG has been one of the most-studied areas in optical physics,[3] in part due to the requirement of nonlinear materials to have a nontrivial second order susceptibility tensor that obey phase matching constraints. Increased SHG response is usually achieved through carefully phase matching several domains of a nonlinear non-centrosymmetric crystal or increasing either the electric field or effective material nonlinearity mainly through enhanced light-matter interaction.[4] Examples of SHG phase matching techniques include birefringent crystals,[5,6] Bragg gratings,[7] quasi-phase matching schemes,[8] and, more recently, quasi-phase-matched photonic crystal slabs[9] and epsilon-near-zero metamaterials.[10] In general, increasing the nonlinearity within a material has recently been accomplished with liquid crystals, photonic crystals, and metasurfaces and metamaterials.[11–13] Finally, SHG can be increased when the local electric field strength is enhanced, causing increased light-matter interactions, as has been shown in waveguides,[14–16] optical microcavities,[17] and membrane metasurfaces.[18]

More specifically, metamaterials offer unique methods of engineering the nonlinear material response that can be leveraged for SHG improvement mainly due to the creation of strong resonant fields, with some relevant examples being magnetic metamaterials,[2] periodically nanostructured metal films,[19] meta-crystals,[20] non-centrosymmetric shaped metallic nanoapertures,[21] electrically controlled plasmonic gratings,[22] and negative-index metamaterials.[23] The emerging field of thin metamaterials fabricated in two dimensions, a.k.a., metasurfaces, enables a new degree of spatial variation of structural parameters across the transverse plane, in addition to the inherently compact footprint, offering precise control over various spatial attributes (e.g., phase, polarization, amplitude, and wavefront) of the emitted nonlinear signal. Recent developments such as twistronic anisotropic configurations made of stacked gratings promise to broaden the applications of ultrathin nanostructures to more complex optical nonlinear effects like photonic memory for optical information storage and realizing reconfigurable devices that can achieve all-optical switching.[24,25]

The metallic (plasmonic) version of metasurfaces have been used to enhance the electric field by surface plasmon resonances and resulting SHG within magneto plasmonic crystals,[26–28] periodic gold (Au) nanorod arrays,[27] and dolmen-type nanostructures.[29] Very weak SHG from thin films mainly due to induced nonlinear surface currents has also been demonstrated by using thin silver films,[30] rutile titanium dioxide ($TiO_2$) in air,[31] holographic silver gratings,[32] and thin amorphous or crystalline silicon films.[33] However, all of these configurations do not exhibit strong electric fields and face challenges in efficiently boosting the SHG process. Other metasurface configurations have also been explored for SHG mainly based on materials with broken centrosymmetry, including membrane metasurfaces,[20] strained silicon nanopillar arrays,[34] lithium niobate (LN) waveguides,[15,35] etched LN nanostructures,[36] LN metasurfaces,[37] arrays of metallic spheres with multipole resonant coupling,[38] v-grooves in monocrystalline Au surfaces,[39] and LN cavities.[40]

However, the research on the contributions of surface effects in SHG of plasmonic nanostructures is still an ongoing effort.[41] Within the electric dipole approximation, SHG is prohibited in centrosymmetric plasmonic metals, but magnetic or higher-order contributions—specifically electric quadrupole and magnetic dipole interactions—can induce SHG, as have been theoretically demonstrated using various modeling frameworks.[42] These contributions are naturally incorporated in free-electron models that describe SHG originating from the bulk of metals.[43–45] As research has increasingly focused on light–matter interactions at the nanoscale, significant attention has been directed towards understanding SHG in noble-metal nanoparticles and nanostructures.[46,47] Within this framework, several studies have demonstrated that the bulk contribution can dominate the SHG response, particularly in elongated structures such as gold split-ring resonators and T-shaped apertures in a gold film.[48,49] Complementary theoretical and experimental investigations involving three-dimensional gold nanoantennas and spherical nanoparticles have shown that the relative significance of bulk and surface contributions depends sensitively on the specific experimental configuration.[50,51] Other studies have identified the normal component of the surface nonlinearity as the primary contributor to SHG in plasmonic nanoparticles.[47,52,53] In particular, a recent review notes that "the surface SHG becomes more and more predominant as the surface-to-volume ratio increases," further suggesting that for nanoparticles, bulk contributions can often be neglected due to their high surface-to-volume ratio.[54] However, while this reasoning appears intuitive, it has been demonstrated that for spherical nanoparticles, size dependence alone is insufficient to definitively identify the dominant nonlinear source.[50]

In this work, we shed further light on the mechanisms that provide SHG enhancement in plasmonics but now by using nonlinear nanocavity arrays forming plasmonic metasurface absorbers instead of nanoparticles or other mainly scattering configurations.[55,56] Note that a single nanoscatterer forming a sole nanocavity will lead to very weak scattering signal that requires the use of microscopes to be detected even in the linear case. In the current work, the reflected signal from our nanocavity array forming a plasmonic metasurface is strong and the second order nonlinear effects are substantially boosted in reflection. More specifically, we experimentally demonstrate $10^5$-fold SHG rate enhancement factors despite that the dielectric material in the nanogap is centrosymmeric with extremely weak second order nonlinear properties. We theoretically investigate the induced nonlinear polarizabilities from both dielectric and metallic parts of the realized metasurface nanocavities and conclude that the strong SHG signal originates primarily from the induced nonlinear surface current distributions along metal. Such nonlinear surface currents are substantially boosted due to the strong field confinement achieved by the current metasurface nanocavity array design. The presented strong SHG due to an ultrathin plasmonic metasurface configuration will lead to new nonlinear sensing and imaging applications.

## 2. Theoretical Section

We fabricate, experimentally measure, and theoretically investigate an ultrathin plasmonic metasurface design. It is made of periodic gold nanostripes grown over a gold mirror and an ultrathin $TiO_2$ dielectric layer. More specifically, the realized plasmonic absorber configuration leads to the formation of an array of nonlinear nanocavities that substantially enhance the local electric field within their nanogaps composed of 4 nm thick $TiO_2$ dielectric layer that is deposited over the gold opaque substrate. The enhanced electric fields inside the nanogaps region increase the induced surface currents along the metallic components, resulting in a significant SHG nonlinear response from the metasurface. Because $TiO_2$ has an extremely weak second order nonlinear susceptibility, the origin of the measured strong second order nonlinear effects observed in this study are due to the gold parts of the metasurface. This is a significant distinction from previous works that use relevant plasmonic metasurfaces solely to increase the electric field in a separate dielectric material with high and usually third order nonlinear susceptibility.[57–59]

The plasmonic metasurface design and its dimensions are shown in Figure 1. Periodic gold nanostripes are positioned over a thin layer of TiO₂ deposited on an opaque and thick underlying Au substrate. The height, width, and periodicity of the nanostripes are $h = 25\ nm$, $w = 90\ nm$, and $p = 145\ nm$, respectively. The intermediate TiO₂ layer is equal to 4 nm thick, while the bottom Au substrate layer is fixed to 100 nm, i.e., significantly thicker than the skin depth of metals to ensure no transmission of either pump or SHG through the substrate. The dimensions of the metasurface and oxide nanogap layer are determined by numerical optimization to find the highest electric field enhancement and subsequently boosted SHG efficiency in our linear and nonlinear simulations, respectively. Then, the metasurface is fabricated and its linear and nonlinear responses are verified experimentally. The performed theoretical modeling is strictly adhered to the fabrication process. Hence, the dimensions and material properties used in the optimal simulation match the values of the fabricated sample obtained by scanning electron microscopy (SEM) images and energy dispersive X-ray spectroscopy data with more details presented in the Supplementary Document. The presented metasurface is excited by an incident plane wave normal to the surface that is linearly polarized with the electric field in the x-direction (transverse magnetic or TM polarization). The other orthogonal transverse electric (TE) polarization does not excite the metasurface resonance and is omitted from this study.[58] However, the metasurface can work for both polarizations if nanopatches or nanocubes are used instead of the current elongated gold nanostripes.[60]

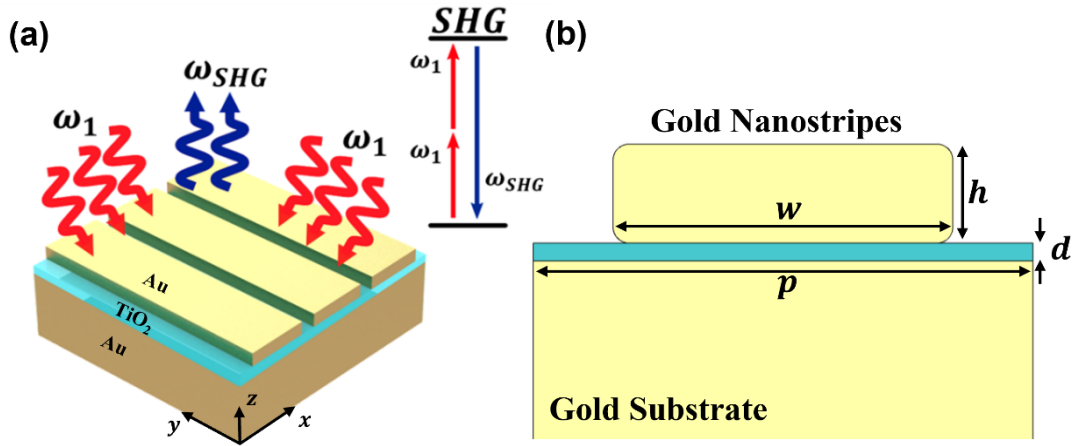

**Figure 1.** (a) Schematic of the plasmonic metasurface nanocavity that achieves enhanced SHG process with generated second harmonic waves in reflection. (b) Dimensions of the metasurface unit cell.

The nonlinear wave equation that describes the SHG process and used in our nonlinear simulations is derived from Maxwell's equations to be:

$$\nabla \times (\mu_r^{-1} \nabla \times \mathbf{E}) - \varepsilon_r k_0^2 \mathbf{E} = \mu_0 \omega^2 \mathbf{P}^{NL}, \tag{1}$$

where $\mathbf{P}^{NL}$ is the combined (bulk plus surface) induced nonlinear polarizability term and the expression $\mu_0 \omega^2 \mathbf{P}^{NL}$ forms a nonlinear source due to the SHG process. This term can be conveniently expressed as a weak form contribution in finite element method solvers, such as COMSOL that is used in our current study. To determine the nonlinear polarization term at the metallic surfaces, we formulate their induced normal surface currents following derivations from previous works related to metallic nanoparticles[50] and not nanocavities like our current work. Hence, we express the normal polarizability term $\boldsymbol{P}_{\perp\perp\perp}^{NL}$ as an equivalent magnetic current density equal to $\boldsymbol{J}_{m,s}^{NL} = \mathbf{e}_\perp \times (\nabla_\parallel P_{\perp\perp\perp}^{NL})/\varepsilon'$, where $\mathbf{e}_\perp$ is the unit vector normal to the surface and $\varepsilon'$ is the permittivity of the dielectric in the nanogaps.[61] All sources of surface polarizability can then be expressed as a set of weak form contributions.

The intensity of the incident wave is fixed to $100\ MW/cm^2$ for the SHG simulations, a value that corresponds to the peak power delivered by our femtosecond laser used in the experimental measurements. A pulsed laser is used to achieve high intensity combined with improved heat dissipation. The undepleted pump approximation is adopted in all of our simulations since the SHG generated wave due to surface effects has lower intensity compared to the used pump. The expression for the nonlinear second-harmonic bulk polarizability, accounting for the nonlocal bulk field contributions induced in isotropic and centrosymmetric materials, including metals, is given below:[50,62]

$$\mathbf{P}_{\text{bulk}}(\mathbf{r}, 2\omega) = \gamma_{bulk} \nabla[\mathbf{E}(\mathbf{r}, \omega) \cdot \mathbf{E}(\mathbf{r}, \omega)], \tag{2}$$

where $\gamma_{bulk}$ is the bulk nonlinear susceptibility characterizing the multipole interactions occurring within the metallic materials and $\mathbf{E}$ is the electric field at the fundamental pump frequency. The nonlinear surface polarizability and its contributions to the total SHG response are crucial to this study, and can be decomposed to normal and parallel components ($\mathbf{P}_\perp$ and $\mathbf{P}_\parallel$ respectively):[50]

$$\begin{aligned}\mathbf{P}_{\text{surface}}(\mathbf{r}^+, 2\omega) &= \mathbf{P}_\perp(\mathbf{r}^+, 2\omega) + \mathbf{P}_\parallel(\mathbf{r}^+, 2\omega) \\ &= \chi_{\perp\perp\perp} \mathrm{E}_\perp(\mathbf{r}^-, \omega) \mathrm{E}_\perp(\mathbf{r}^-, \omega) \mathbf{e}_\perp \\ &\quad + \chi_{\parallel\parallel\perp} \mathrm{E}_\parallel(\mathbf{r}^-, \omega) \mathrm{E}_\perp(\mathbf{r}^-, \omega) \mathbf{e}_\parallel,\end{aligned} \tag{3}$$

where ⊥ and || indicate the components perpendicular and parallel to the surface, with corresponding unit vectors $\mathbf{e}_\perp$ and $\mathbf{e}_{||}$, respectively. Since these surface components are always computed along the metal-dielectric interfaces, the plus and minus sign superscripts denote the dielectric and metallic portion of the metasurface, respectively. For isotropic and centrosymmetric materials, the perpendicular and parallel components are mapped to the susceptibility tensor components following the coordinate system shown in Figure 1 as: $\chi_{\perp\perp\perp} = \chi^S_{zzz}$, $\chi_{\perp||\,||} = \chi^S_{zxx} = \chi^S_{zyy}$, $\chi_{||\,||\perp} = \chi^S_{xxz} = \chi^S_{xzx} = \chi^S_{yyz} = \chi^S_{yzy}$.[63] While $\chi_{\perp\perp\perp}$, $\chi_{||\,||\perp}$, and $\gamma_{bulk}$ can be understood in terms of Rudnick and Stern parameters,[30,64] we instead use more accurate experimental values measured in the literature for gold surfaces to be: $\chi_{\perp\perp\perp} = 1.35 \times 10^{-18}\ m^2V^{-1}$, $\chi_{||\,||\perp} = 3.86 \times 10^{-20}\ m^2V^{-1}$, and $\gamma_{bulk} = 6.06 \times 10^{-21}\ m^2V^{-1}$.[65] These experimental values provide a very good match between simulations and measured results, as will be shown later in the results section. Note that while these values are very low, they must be integrated over a nanometer scale surface making them comparable or even larger in our case compared to bulk susceptibilities. For homogeneous and isotropic materials like those used here, the remaining contributions to bulk polarization are zero since the divergence of the electric field is zero. Additionally, the surface polarization term $\chi_{\perp||\,||}E_{||}(\mathbf{r}^-,\omega)E_{||}(\mathbf{r}^-,\omega)$ is negligible because $\chi_{\perp||\,||}$ is significantly smaller than the terms $\chi_{\perp\perp\perp}$ and $\gamma_{bulk}$, hence, it is ignored in our current modeling efforts.[63]

We now combine the normal components of Equations (2) and (3), noting that the perpendicular polarizability terms are a linear combination of bulk and surface parts, where the bulk can be expressed as an effective surface current contribution:[30,66]

$$P_\perp(\mathbf{r}^+, 2\omega) = \left(\frac{\gamma_{bulk}}{\varepsilon(2\omega)} + \chi_{\perp||\,||}\right)\mathbf{E}(\mathbf{r}^-,\omega)\cdot\mathbf{E}(\mathbf{r}^-,\omega)$$
$$+ (\chi_{\perp\perp\perp} - \chi_{\perp||\,||})E_\perp(\mathbf{r}^-,\omega)E_\perp(\mathbf{r}^-,\omega). \quad (4)$$

Again, since $\chi_{\perp||\,||}$ is orders or magnitude smaller than the other coefficients, the polarizability expression is simplified to:

$$P_\perp(\mathbf{r}^+, 2\omega) = \frac{\gamma_{bulk}}{\varepsilon(2\omega)}\mathbf{E}(\mathbf{r}^-,\omega)\cdot\mathbf{E}(\mathbf{r}^-,\omega) + \chi_{\perp\perp\perp}E_\perp(\mathbf{r}^-,\omega)E_\perp(\mathbf{r}^-,\omega). \quad (5)$$

Next, we consider experimental constraints to further abridge the aforementioned polarizability term. For the metasurface under normal incidence of linearly polarized light along the x-direction, the localized plasmon resonance observed in the dielectric nanogap layer will result in high electric field, especially in the z-direction. At the interface of the dielectric layer with the underlying gold substrate and the metasurface nanostripes above, the obtained resonance results in $E_\perp$ being the prominent element among the electric field vector components while the electric field in the bulk of the metal is much lower because of attenuation due to the skin depth effect. Hence, Equation (5) can be rewritten as:

$$P_\perp(\mathbf{r}^+, 2\omega) = \chi^*_{\perp\perp\perp} E_\perp(\mathbf{r}^-, \omega) E_\perp(\mathbf{r}^-, \omega), \quad (6)$$

where we define $\chi^*_{\perp\perp\perp} = \left(\chi_{\perp\perp\perp} + \frac{\gamma_{bulk}}{\varepsilon(2\omega)}\right)$ as the effective normal susceptibility due to the metasurface. When the induced surface polarizabilities derived in this section are used in our theoretical modeling, we obtain simulated SHG nonlinear results that accurately match the measured experimental findings, meaning that surface induced nonlinear currents are the dominant mechanism of the presented boosted SHG process. Note that the perpendicular and parallel induced electric field components are comparable along the metal dielectric interface and are depicted in Figure S1 in section 1 of the Supporting Information document.

## 3. Results and Discussion

The metasurface was fabricated and more details of the process are presented in Section 2 of the Supporting Information document. To experimentally determine the optical characteristics of the fabricated metasurface, the fundamental plasmonic resonant mode of the sample is first computed by measuring the reflection of white light upon the sample. Absorption of light occurs at the fundamental and higher order resonances of the metasurface, corresponding to resonant wavelengths with high electric field enhancement within the nanogap dielectric layer. The metasurface geometry is optimized to feature near total absorption at the fundamental resonance to foster the highest SHG efficiency but also has strong absorption in the second harmonic wavelength to further boost SHG process. Two experimental setups with similar optical paths are developed to measure the linear reflectance and nonlinear harmonic generation properties of the metasurface, with the reconfiguration between the two setups only requiring the change of a few optical components.

More specifically, the linear reflectance measurements are performed using a broadband white light source (Thorlabs OSL2IR fiber coupled tungsten-halogen light) which is collimated and polarized orthogonal to the direction of the metasurface nanostripes. It is redirected to the metasurface at normal incidence with a broadband 50:50 beamsplitter (BSW29 600-1700 nm) and focused with a 10X (0.26 NA) objective over an illumination area of roughly 40 μm in diameter, sufficiently small compared to the 100μm wide metasurface sample to avoid boundary or edge effects. Normal incidence illumination is used for all experiments and simulations because the metasurface design results in the highest SHG response at normal incident angle. Simulations that compute the reflection and SHG intensity at oblique incident angles are presented and discussed in Section 3 of the Supporting Information. At the resonant wavelengths of the metasurface, the incident light is absorbed, while the remainder of the light spectrum is reflected to a camera that helps to correctly position the metasurface. The broadband light source is spectrally dispersed by a Horiba (iHR550) spectrometer with two Horiba Synapse charge-coupled devices (CCDs) that feature peak quantum efficiency in the visible and infrared (IR), respectively.

Using the optical setup presented in Figure 2(a), the normalized reflectance is computed by capturing the spectra of the white light incident upon the metasurface normalized to the spectra collected from the gold mirror (substrate) background, expressed as $R = (C_{meta} - C_0)/(C_{mirror} - C_0)$, where $C_{meta}$ and $C_{mirror}$ are the counts measured from the metasurface and the mirror-like substrate, respectively, and $C_0$ is the measured dark count background for each CCD sensor. To obtain the entire spectra data over the broad wavelength range of 500-1600nm utilized to characterize the metasurface in the fundamental and SHG regions, both IR and visible CCDs are used and combined as depicted in Figure 2(a). Moreover, the SEM image of the plasmonic metasurface, showing the gold nanostripes deposited over the darker underlying ultrathin $TiO_2$ film is demonstrated in Figure 2(b). We use e-beam lithography to realize the nanostripes and their pattern is repeated over a square area of edge length 100 μm, as described in more detail in section 2 of the Supporting Information document. This technique achieves minimum roughness despite fabricating nanoscale metallic structures.

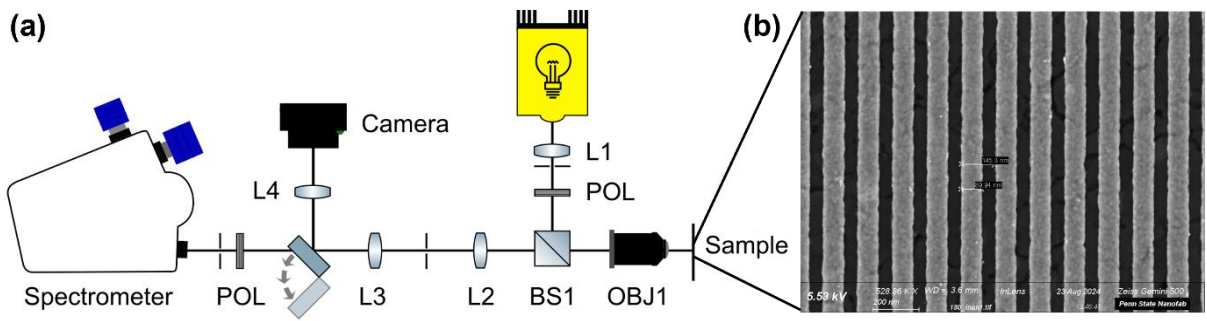

**Figure 2.** (a) Reflectance measurement experimental set-up. (b) SEM image of the fabricated plasmonic metasurface.

The experimentally measured reflectance data agrees very closely with the simulated results, and the comparison is shown in Figure 3. The metasurface nanocavity exhibits a fundamental resonance at 1220 nm, where almost total absorption of incident light is achieved, and a more shallow higher-order resonance at 678 nm. This higher order resonance is sufficiently close to the SHG wavelength leading to further enhancement of SHG efficiency, as will be shown in the nonlinear results presented later in the paper. In our work, we use the fundamental resonance to absorb the incident light leading to a large field enhancement within the dielectric layer located at the nanogaps of the nanocavity array. The higher-order resonance is detuned on purpose because we need a more reflective or less absorptive resonant response at the SHG wavelength (black dashed line in Figure 3) but combined with strong field enhancement in the nanogaps. The field profiles at fundamental and SH wavelengths along with the induced surface currents at the SH wavelength are plotted and discussed in more detail in the Supporting Information Section 4.

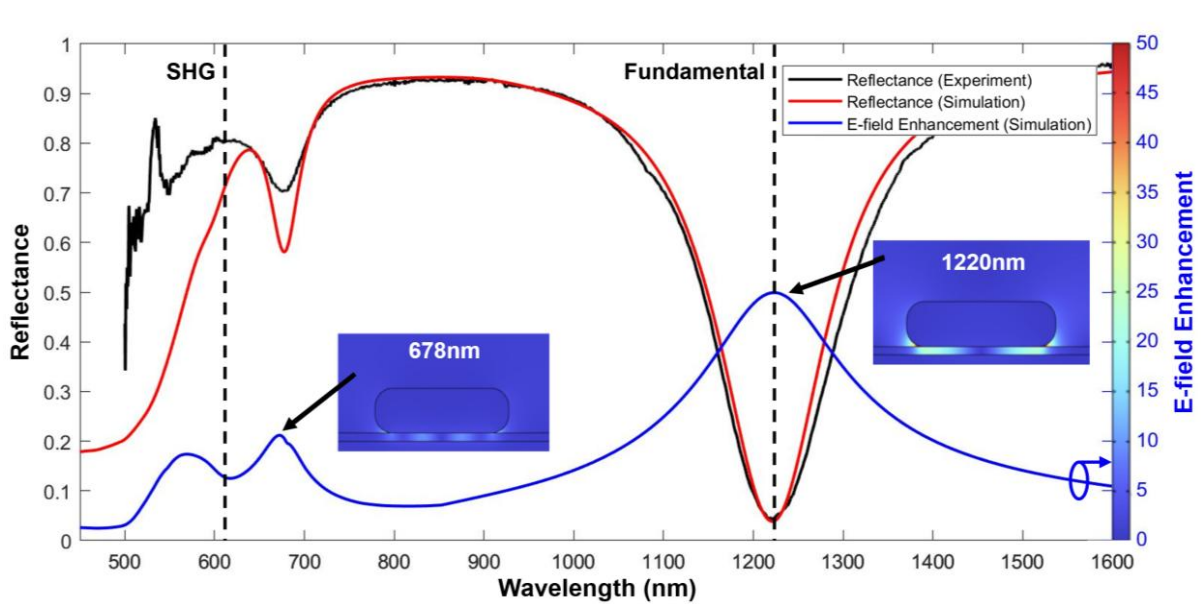

**Figure 3.** Measured (experiment) and simulation linear reflectance results along with the numerically simulated electric field enhancement in the fundamental and higher-order resonance, where strong electric field enhancement is obtained at the nanogap. The blue line (electric field enhancement) has values shown in the right y-axis.

The second harmonic generation of the metasurface was measured by adapting the optical path of the reflectance measurements to include a higher intensity pulsed laser source instead of the low intensity white light, using the setup depicted in Figure 4(a). Specifically, a femtosecond pulsed laser and optical parametric oscillator (OPO) system (Coherent Chameleon OPO-Vis, 80 MHz, 140 fs) is used to achieve the high intensity required to trigger SHG and limit the thermal degradation of the sample, while being tunable over a range of wavelengths including the fundamental resonance. In particular, we use a femtosecond pulsed laser because it can provide high laser intensities combined with ultrafast duration that are critical to measure the usually weak optical nonlinear effects. This happens without degrading the sample due to accumulated heat which usually occurs when using high incident intensities generated either by continuous wave lasers or high repetition rate and broader pulse duration (pico/nanosecond) lasers.[67] The laser power directly from the OPO requires attenuation, which is done through a half-wave plate and a polarized beamsplitter. A 900 nm long-pass filter is then used to ensure that any remainder of the 780 nm pump is removed from the optical path, and a wire grid polarizer is used to polarize the incident laser pulse. An 800 nm short-pass filter redirects the reflected OPO signal into a 50X (0.42 NA) objective lens to achieve a small spot size (roughly 8 μm in diameter) combined with sufficient intensity to realize a strong SHG signal emitted from the metasurface. The SHG signal propagates opposite of the pump, being collimated through the objective. The fundamental pump wavelength is removed with a short pass filter, before reaching the spectrometer for measurement.

The SHG signal is detected for a range of pump wavelengths that span the width of the fundamental resonance, while SHG signals are obtained at half wavelength values, as expected, with results shown in Figure 4(b). Each curve on the bottom axis depicts a 1-second collection of the SHG with the pump wavelength set at 6.5 mW on the same area of the metasurface. The 1-second period of data acquisition was determined for its convenience while staying below the saturation limit of the spectrometer used to collect data. The largest response is obtained at the SHG wavelength of 610 nm, which corresponds to a pump wavelength of 1220 nm, and is exactly at the center of the metasurface fundamental resonance where the maximum field enhancement in the nanogap is obtained, as was shown before in Figure 3. When the wavelength is tuned farther from this resonance, the total SHG signal is reduced until it is undetectable where the linear absorption reaches zero. Such effect confirms that the enhanced fields within the dielectric layer under high absorption increase the net second-order polarizability of the nanostructure. This is notable since both $TiO_2$ and gold have no significant second-order susceptibility in the bulk, so the SHG is a consequence of the surface effects along the gold features of the nanostripes and substrate, amplified significantly by the field enhancement induced in the nanogap due to the metasurface resonance.

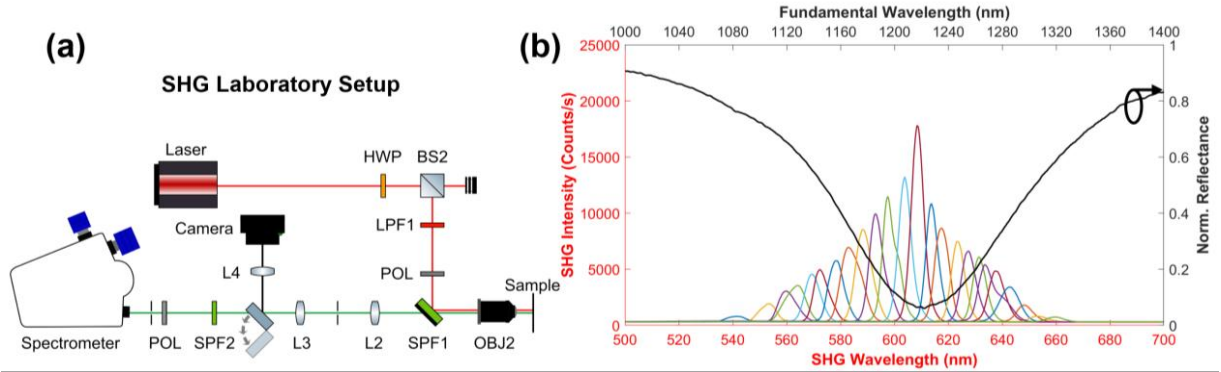

**Figure 4.** (a) Nonlinear SHG experimental setup. (b) Measured SHG spectrum for a range of pump wavelengths spanning half values of the fundamental metasurface resonance. The black line in (b) is the measured linear reflectance with values shown in the right y-axis and wavelength spectrum depicted in the upper x-axis.

To further substantiate the metasurface as the SHG source, we experimentally and theoretically investigate the SHG potential of the bare $TiO_2$ film and gold mirror substrate layer beneath, i.e., no metasurface. The experimental and theoretical simulation results presented in Figure 5(a) match with great accuracy. For the experimental data, the area of illumination is shifted off the metasurface region and the spectra is captured. To ensure the optical path and focus on the surface is maintained, the metasurface alignment is verified by capturing a uniform intensity of SHG signal across the metasurface area with 8 mW of incident pump laser signal. When the area of illumination is shifted slightly off the metasurface nanostripes area, the SHG signal disappears in the noise until the pump power is increased to a higher value of 36 mW. This peak cannot be compared with the SHG signal from the metasurface, which is orders of magnitude larger, but is plotted in the inset of Figure 5(a), reaching only approximately 25 counts per second higher than the noise floor. Increasing the intensity above 36mW increases this peak, but at intensities over 50mW thermal damage occurs to the illuminated area and its surroundings. No thermal damage was detected below the 50mW value since there was no degradation of the measured SHG response signal or by visual inspection of the sample using SEM images. Moreover, the sample has been exposed to open air for most of its life and is robust not only to laser incident power but also laboratory environmental degradation. Fine tuning of the z-direction was attempted to verify that no compensation was needed due to the height of the metasurface or any small displacement due to the movement, but with the objective's depth of field at 1.6 μm, these were found to be inconsequential to the low SHG intensity of the bare (no nanocavity) surface. Simulations in Figure 5(a) are constructed for a distribution of incident powers, which are accurately mapped to the Gaussian profile of a transform limited 140 fs pulse to compare with the experimental data.

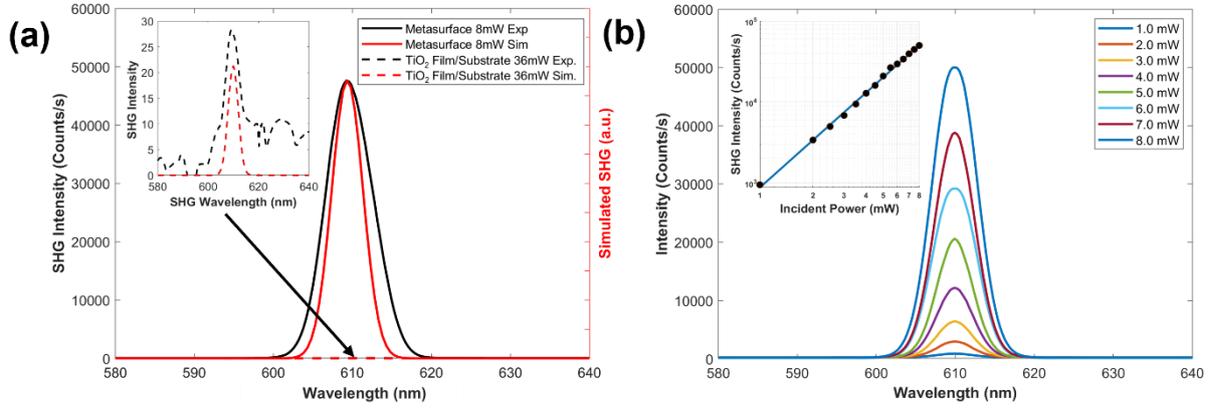

**Figure 5.** (a) Measured and simulated SHG signal from the metasurface using 8 mW input laser power compared to the signal off the metasurface nanostripes (i.e., incident laser impinging on the bare $TiO_2$ layer on gold substrate) using 36 mW laser power. Even with significant boost in pump intensity, the results of the bare (no nanocavity) surface are very low, as shown in the inset zoomed in plot. (b) Measured SHG intensity of the metasurface captured for a range of incident pump laser powers. The SHG intensity versus incident power has a second order relationship with slope 1.97, as depicted in the inset.

To accurately compare the response of the laser with the simulations, we consider a Gaussian beam, as mentioned before, and iterate over its constitutive wavelengths with their relative intensities. Since the spectrometer collects data in units of counts per second, a linear transfer function was used to equate the simulations with the peak counts of the spectrometer. For this reason, the SHG intensity obtained by simulations (red solid line in Figure 5(a)) uses arbitrary units. Next, the metasurface nanostripes are removed from the simulations and the results are scaled using the previously used transfer function leading to a small peak with an intensity of approximately 20 counts per second, which compares well to the collected experimental data. We conclude that both simulation and experimental results clearly confirm that the strong SHG signal obtained originates from the high field enhancement caused by the metasurface nanocavity under resonant excitation. This field enhancement amplifies the SHG process along the gold surfaces of the metasurface and underlying gold substrate, creating an overall effect significantly higher than can be explained from effective bulk nonlinearities in the $TiO_2$ nanolayer. In Figure 5(b), the measured SHG spectra is shown for a range of pump laser intensities, with the peak intensity plotted against the incident laser power in the inset. The slope of this line is determined to be 1.97 with a high-quality fit ($R^2 = .9982$), which verifies the expected second order power dependence, a characteristic behavior of the SHG process. Note that the SHG efficiency can be further increased if we use higher incident power values. However, here it is kept relatively low to not reach saturation effects or thermal damage.

Note that $TiO_2$ is a centrosymmetric material with negligible bulk and surface second order nonlinear susceptibilities. Next, we confirm by performing extensive nonlinear simulations that the surface nonlinear susceptibility of metal, i.e., not bulk nonlinearity from $TiO_2$, is responsible for the strong SHG response observed in our experiments. To prove this point we allow the $TiO_2$ layer to have a reasonable and non-zero bulk nonlinear susceptibility and show that the SHG intensity computed by simulations is insignificant compared to similar nonlinear simulations that include the surface nonlinear properties of metal. To represent the highest reasonable bulk nonlinear susceptibility that could be attributed to the $TiO_2$ ultrathin layer, since there is no relevant data in the literature, we consider another related oxide, named silicon oxide ($SiO_2$), that has measured values of bulk nonlinear susceptibility up to $d_{11} = 0.3\ pm/V$.[3,68,69] However, it should be noted that such values refer to highly structured quartz ($SiO_2$) crystals, which have higher potential to realize noticeable second order nonlinearities than the currently used amorphous oxide variants, similar to the $TiO_2$ ultrathin layer. Indeed, the 4 nm $TiO_2$ thin film used in our metasurface design is best described as amorphous due to used atomic layer deposition growth process, as described in Section 2 of Supporting Information.

More specifically, to rule out the possibility of such a bulk nonlinear susceptibility in the oxide layer as a potential contribution to the observed SHG signal, we compare in Figure 6 the simulation results based on nonlinear surface effects (similar to theoretical results presented before and shown by red line in Figure 6) to an identical in terms of geometry simulation model but when removing all nonlinear surface contributions and instead using a bulk effective nonlinear susceptibility of $d_{eff} = 0.3\ pm/V$ for the dielectric material (blue line in Figure 6) which is the highest measured value for structured $SiO_2$ crystals.[3,68,69] Note that this nonlinear susceptibility term is aligned with the incident electric field along the x-direction to make sure that maximum SHG enhancement can be realized. Even in this case, the bulk nonlinear simulations (no surface effects, blue line in Figure 6) result in very weak SHG, with a peak intensity of 191 counts per second, approximately 260 times smaller than the correct simulations with the full nonlinear surface effects present (red line in Figure 6) that were used to verify the obtained experimental results. Therefore, Figure 6 further proves that the measured enhanced SHG is due to surface nonlinear effects along the metal dielectric interface of the nanogap.

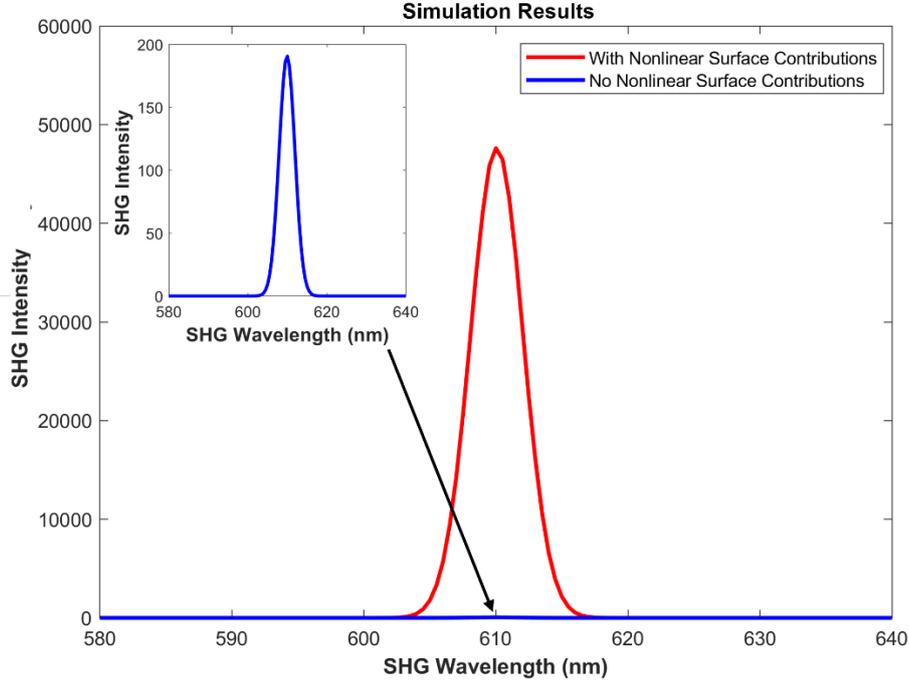

**Figure 6.** SHG simulation results of the plasmonic metasurface design when the induced nonlinear surface contributions to the second order polarizability are taken into account (red line) compared to simulations without these surface contributions (blue line). The latter results in very weak SHG process compared to simulations based on full nonlinear surface effects that match the experimental results. In both simulations, the underlying dielectric has a bulk nonlinear susceptibility of $d_{eff} = 0.3\ pm/V$ to demonstrate the scale of the surface current response over any possible non-zero bulk nonlinearities.

Finally, by measuring the SHG emitted power experimentally in our work, we compute the SHG enhancement factor defined as the ratio of the SHG power of the nanocavity array to the SHG power from the underlying TiO$_2$ thin film and gold substrate, i.e., design without nanocavities, by using the formula: $EF = P_{cavity}/P_{TiO_2}$. Here, we use the measured counts per second depicted in Figure 5 to represent these powers. The nanocavity array produced $P_{cavity} = 48k$ counts per second due to SHG measured by the detector at 8 mW, while the bare TiO$_2$ surface without nanocavities resulted in 20 counts per second with the incident power set to the higher value of 36 mW. Projecting the number of bare surface counts to 8 mW results to a value of $P_{TiO_2} = 4.4$ counts per second. This gives an SHG enhancement factor of $EF = P_{cavity}/P_{TiO_2} = 1.1 \times 10^5$. Such experimentally measured strong SHG enhancement is uncommon for second order nonlinear processes and stems from the metallic surface nonlinear effects, as theoretically explained before, which is in direct contrast to previously used bulk dielectric nanofilms made of indium-tin-oxide or other materials exhibiting epsilon-near-zero response.[70-72]

## 5. Conclusions

The presented work demonstrates experimentally an optimized ultrathin nanocavity array configuration forming a plasmonic metasurface absorber that achieves strong SHG signal using only centrosymmetric materials. While such materials have very low second order nonlinear susceptibility, it is proven that the observed enhanced nonlinear SHG process can be attributed to the surface nonlinear susceptibility of the metal aided by the significant electric field enhancement that occurs at the nanogaps due to the formed nanocavity resonance. Rigorous theoretical nonlinear simulations are performed to provide further insights and explain the SHG results that consider both bulk and surface linear and nonlinear material properties. It is found that the induced nonlinear surface currents along the metal-dielectric interfaces between nanostripes, dielectric, and substrate are the main reason of the obtained $10^5$-fold boosted SHG process achieved by the currently demonstrated metasurface nanocavity array design. Hence, we not only present tangible experimental results that demonstrate substantially enhanced SHG process stemming solely from surface nonlinearities and not bulk non-centrosymmetric materials, but we employ advanced computational models to clearly explain the complex underlying physical mechanisms behind this unusual effect. The presented strong harmonic generation response from an ultrathin plasmonic metasurface is envisioned to lead to new nonlinear sensing and imaging applications.

**Acknowledgements**

This work was partially supported by the National Science Foundation (NSF) DMR-2224456 and the Penn State Applied Research Laboratory Walker Fellowship.

# Supporting Information

# Leveraging Plasmonic Nanocavity Arrays Forming Metasurfaces to Boost Second Harmonic Generation due to Surface Effects

Sky Semone,[1,2] Matthew J. Brandsema,[2] Thang Hoang[3], and Christos Argyropoulos[1,*]

[1]Department of Electrical Engineering, Pennsylvania State University, University Park, PA, 16803 USA
[2]Applied Research Labs, Pennsylvania State University, University Park, PA, 16803 USA
[3]Department of Physics and Material Science, University of Memphis, Memphis, TN, 38152 USA
*cfa5361@psu.edu

Keywords: metasurfaces, nanocavities, nonlinear optics, second harmonic generation

1. **Metasurface Induced Field Distributions at Fundamental Wavelength**

In the simulation model derived in our work, we assume that the normal contribution of the nonlinear polarizability is the primary contribution to the observed SHG signal, while the parallel component can be ignored due to much lower value. Equation (4) in the main text gives the formula for the second-order polarizability before these simplifications are made, where the term $\chi_{\perp\parallel\parallel}\mathbf{E}(\mathbf{r}^-,\omega)\cdot\mathbf{E}(\mathbf{r}^-,\omega)$ can still be comparatively large if the parallel components of the electric field within the gold components $\mathbf{E}(\mathbf{r}^-,\omega)$ are substantial compared to the normal components. However, analysis of the electric field components proves that this is not the case, and the normal and parallel components are approximately equal in amplitude across the gold/dielectric interfaces, as seen in Figure S1.

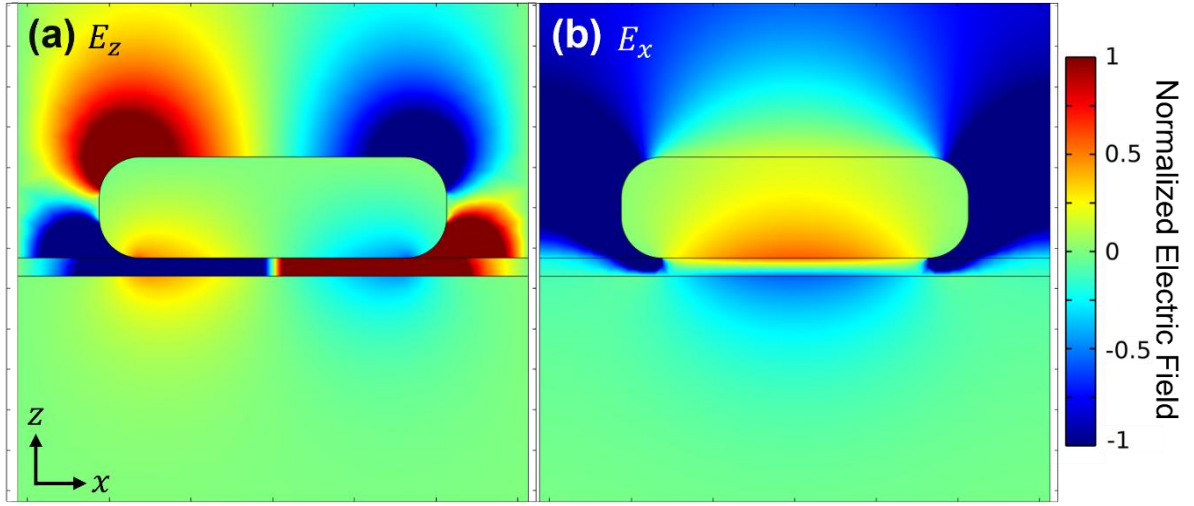

**Figure S1.** Normalized real electric field values induced in the (a) z- and (b) x-direction and plotted along a cross-section of the metasurface.

Since the electric field values are approximately the same for the normal and parallel components at the interface, we use the perpendicular and parallel surface susceptibility tensor elements ($\chi_{\perp\perp\perp} = 1.35 \times 10^{-18}\ m^2V^{-1}$, $\chi_{\parallel\parallel\perp} = 3.86 \times 10^{-20}\ m^2V^{-1}$) to determine that the parallel contributions to the total SHG are much weaker than the normal. To further verify this claim, the parallel contributions to the second order polarizability are expressed in the primary SHG COMSOL model used in the main text. In this case, the sum contributions of these parallel terms are 289 times less than the contributions of the normal expressions.

2. **Metasurface Fabrication Details**

The metasurface dimensions are determined by optimizing our simulation models to achieve the strongest SHG response within reasonable fabrication constraints. A portion of a 100 mm prime grade, [100], P-type wafer is used as the substrate for all subsequent steps. To achieve strong adhesion of the following layers to the substrate, a Temescal FC-2000 electron beam evaporator is used to deposit a 10 nm bonding later of chromium to the substrate. Next, the same machine is used to deposit 100 nm of gold (Au) at 2 A/sec with the goal to form the underlying substrate of the metasurface.

Atomic layer deposition (ALD) of 4nm thick $TiO_2$ is performed with the Lesker ALD LE 150 using 90 cycles of TDMAT precursor at 100°C. The final thickness of the deposition is confirmed by a Woolam M-2000 ellipsometer using reference measurements before and after the $TiO_2$ deposition. The refractive index of $TiO_2$ is also determined from this measurement to be 2.1, which is typical value for amorphous thin films grown by ALD.[1] This is essential for translating the COMSOL model to the fabricated device, since the optical response is highly dependent upon the index of refraction within this layer where the maximum field enhancement is obtained due to the nanogap.

A grating pattern of 90 nm wide with 145 nm periodicity is repeated over a square area of length 200 um using KLayout mask layout software. ZEP 520A is selected for the pattern resist because of its high sensitivity to the small pitch of the metasurface. The resist is spun on the sample wafer and baked on a hot plate for 180 seconds. A Raith EBPG 5200 direct-write electron beam pattern generator is used to perform lithography of the grating using a 1 nm step size. A dosing of 180 uC/cm^2 is determined to give best edge resolution and most accurate size through a series of dosing tests conducted before final fabrication. The resist is then developed in amyl acetate and quickly rinsed in IPA for 15 s. Finally, $O_2$ plasma cleaning with the Templa M4L removes any debris or residue from the lithography process and provides the clean edges of the grating in addition to improving adhesion of the metal to the oxide surface.

An aluminum shadow mask is affixed to allow deposition only in the area near the grating to increase the speed and quality of the final liftoff. A 25 nm layer of gold is then deposited with the Temescal FC-2000, by using a 10 RPM rotation and slow deposition rate of 0.5 A/s with the goal to provide a uniform layer thickness. Special care is taken to ensure the temperature of the substrate does not rise far above ambient temperature to ensure the quality of the resist features. Liftoff is completed by placing the sample upside-down in heated PRS-3000 remover for 2 hours. Flushing the surface with clean, heated PRS-3000 aids with the complete liftoff of the entire metasurface area. A final clean in acetone and IPA removes the PRS-3000 and concludes the sample fabrication process.

Interestingly, in this study, the nanogap thickness formed between the nanostripes and the gold substrate layer proved critical to obtaining high electric field enhancement and boosted SHG. The thickness of $TiO_2$ is first verified with a Woolam M-2000 ellipsometer following the deposition of the thin layer. The process of e-beam lithography requires use of high energy electrons to expose the ZEP-520a positive resist, where the metasurface is present in the final design. To ensure subsequent processes do not deteriorate the $TiO_2$ layer, energy dispersive spectroscopy (EDS) is performed using UltimMax 100 SSD Detector to determine the elemental composition of a small area under the metasurface, as depicted in the SEM image shown in Figure S2(a). The EDS results are presented in Figure S2(b), which validate the $TiO_2$ nanogap layer due to the existence of both constitutive elements in the chemical analysis. All other elements are also accounted for in these EDS results, such as silicon from the substrate, carbon residue from the lithography resist and liftoff process, and aluminum from the bonding layer between the silicon substrate and gold substrate layer, acting as mirror.

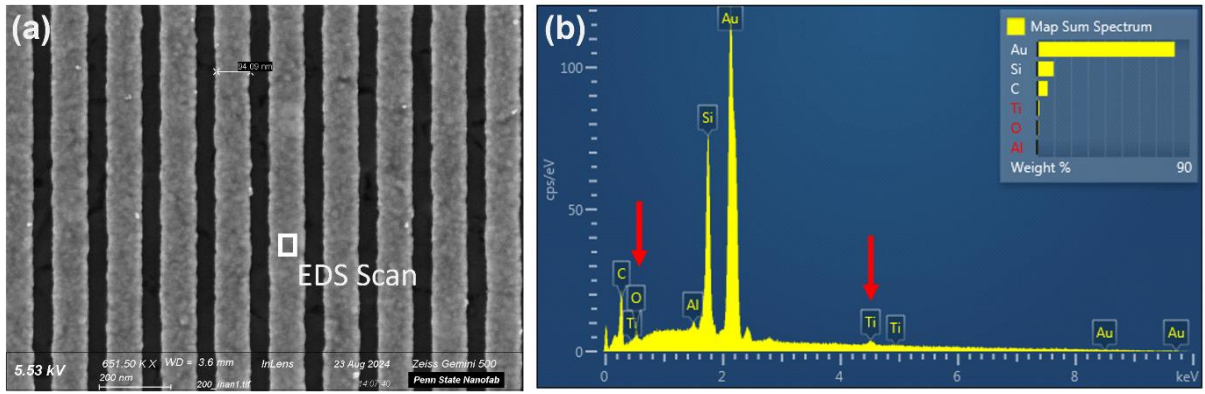

**Figure S2.** (a) EDS is performed over a small area confined entirely over the metasurface nanostripes to verify the existence of $TiO_2$ under them. (b) Titanium is determined to be present in the region probed, which can only be traced to the $TiO_2$ nanogap layer deposited under the gold nanostripes.

### 3. Oblique Incidence Illumination

As a source for second harmonic generation, it is beneficial for the nanocavity array to be usable over small variations of incident illumination mainly for ease of alignment purposes. To demonstrate how oblique incidence response affects SHG, the linear reflectance and normalized SHG intensity are computed by simulations over a range of pump wavelengths and incident angles and presented in Figures S3(a) and S3(b), respectively. Even when the incident light (pump) is at $\pm60$-degree angles from normal incidence, the fundamental and secondary linear resonant reflectance dips remain near their optimal minimum values close to zero at 1220 nm and 678 nm, respectively, as can be seen in Figure S3(a). Since the absorption and resulting field enhancement at the fundamental resonance (1220 nm) is mainly responsible for the large increase in second harmonic generation, the computed SHG intensity remains high at the fundamental (pump) wavelength over a deviation of $\pm20$ degrees from normal, as depicted in Figure S3(b). However, the strongest SHG is obtained for pump at normal incidence which is clearly shown in Figure S3(b).

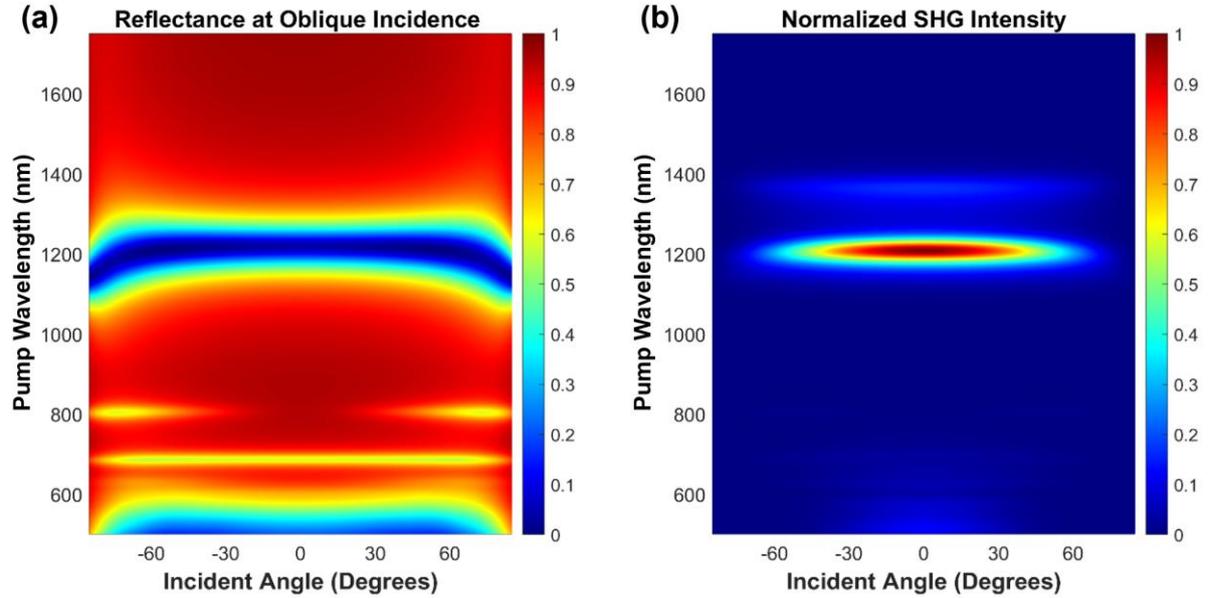

**Figure S3.** (a) Computed linear reflectance for a range of pump wavelengths and oblique incident angles, where zero degrees is normal incidence. The fundamental resonance at 1220 nm and the secondary at 678 nm remain consistent over a range of ±60 degrees from normal incidence, indicating that the nanocavity array is resilient against angular misalignment. (b) Computed normalized SHG intensity as a function of the pump wavelength and oblique incident angle, demonstrating that the nanocavity array forming a metasurface results in high SHG values when excited at the fundamental wavelength (1220 nm). The illumination can be several degrees off from normal incidence without substantially affecting the SHG rate. However, the use of normal incidence in the experimental setup leads to maximum SHG rate.

### 4. Fundamental/Second Harmonic Fields and Nonlinear Surface Currents

The plasmonic metasurface absorber design presented in our work leads to pronounced SHG enhancement due to localized gap plasmon fundamental and higher-order resonant modes formed inside the extremely subwavelength plasmonic nanocavities at $\omega$ (1220 nm) and $2\omega$ (610 nm) frequencies, respectively. The real electric field component distribution profiles at these two frequencies are plotted in Figure S4 and are found to overlap in the nanocavity, especially the dominant $E_z$ component, along several regions of the metal-dielectric interfaces. Note that the fields at SH wavelength (610 nm) are very similar to the higher-order resonance mode field distribution obtained at 678 nm presented in Figure 3 in the main paper. Hence, the resonance at the SH wavelength of 610 nm is still reminiscent of the higher-order resonance, however, with the advantage of stronger reflection (see Figure 3 in main paper), i.e., less absorption, which is beneficial to increase the SHG efficiency.

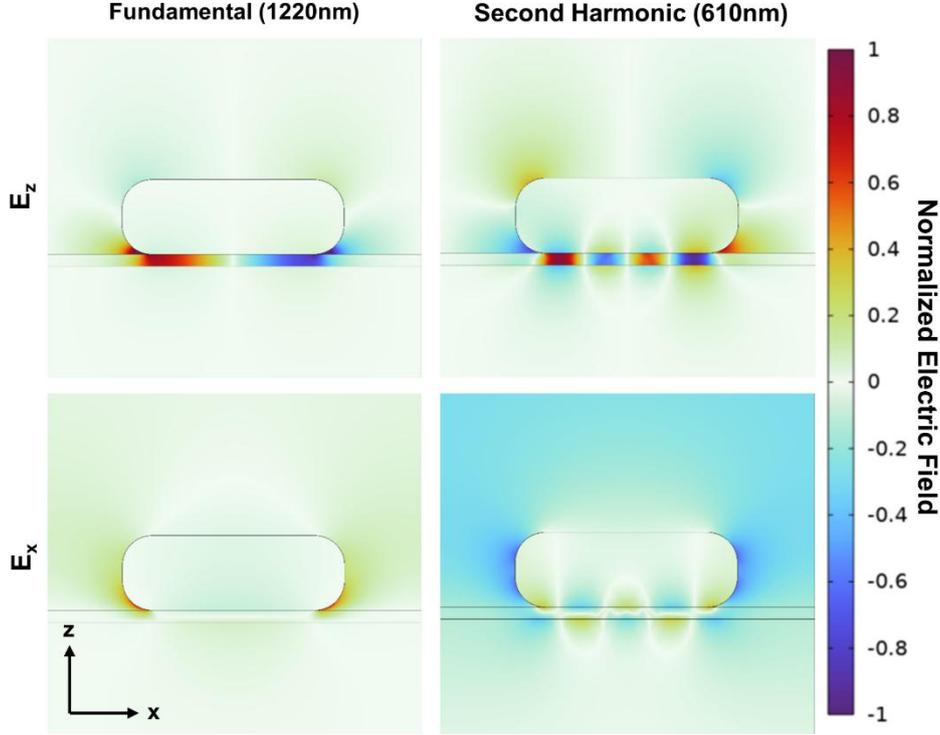

**Figure S4.** The z- and x-components of the real electric field components are computed by simulations and presented at the fundamental 1220 nm (left panels) and second harmonic 610 nm (right panels) wavelength. All results are normalized to the maximum value of the induced electric field within the dielectric layer inside the plasmonic nanocavity.

The normal surface second order susceptibility of gold is the dominant nonlinear component that mainly creates the measured SHG efficiency enhancement presented in our work. The value of this dominant component is found in the literature and given in the main paper. It is repeated here for convenience: $\chi_{\perp\perp\perp} = 1.35 \times 10^{-18}\ m^2V^{-1}$. Since this is the dominant nonlinear component and the bulk second order susceptibilities of the materials used are much weaker, the induced at SH wavelength normal polarizability term $P^{NL}_{\perp\perp\perp}(2\omega)$, which is analogous to $\chi_{\perp\perp\perp}$ and normal electric field at fundamental wavelength (see Eq. (6) in the main paper for exact formula), will create an equivalent normal surface current density given by the formula (repeated here from main paper): $J^{NL}_s(2\omega) = \mathbf{e}_\perp \times \left(\nabla_\parallel P^{NL}_{\perp\perp\perp}(2\omega)\right)/\varepsilon'$, where $\mathbf{e}_\perp$ is the unit vector normal to the surface and $\varepsilon'$ is the permittivity of the dielectric in the nanogaps. Consequently, this equivalent normal surface current density will create electric fields $E(2\omega)$ at SH wavelength boosted only in the case of nanocavities that will lead to the measured strong SHG efficiency.

More specifically, the boosted normal surface current density at SH wavelength due to the plasmonic nanocavity is computed and plotted as contour distribution in Figure S5(a). The same surface current density is plotted in Figure S5(b) when the gold nanostripe is missing, meaning there is no nanocavity. The values of the latter surface current (Figure S5(b)) are substantially weaker compared to the nonlinear current boosted due to the nanocavity effect (see Figure S5(a)), since there is no cavity resonance effect. Moreover, the generated electric field at SH wavelength due to the surface current is orders of magnitude stronger in the case of nanocavity (Figure S5(a)) compared to without the nanocavity (Figure S5(b)). This is the main reason for the enhanced SHG efficiency due to the nanocavity array measured experimentally in our work and verified with nonlinear simulations in the main paper. Hence, here, it is explicitly proven why the SHG is substantially boosted due to surface nonlinearities prevailing in the nanocavity array configuration forming a plasmonic metasurface absorber design.

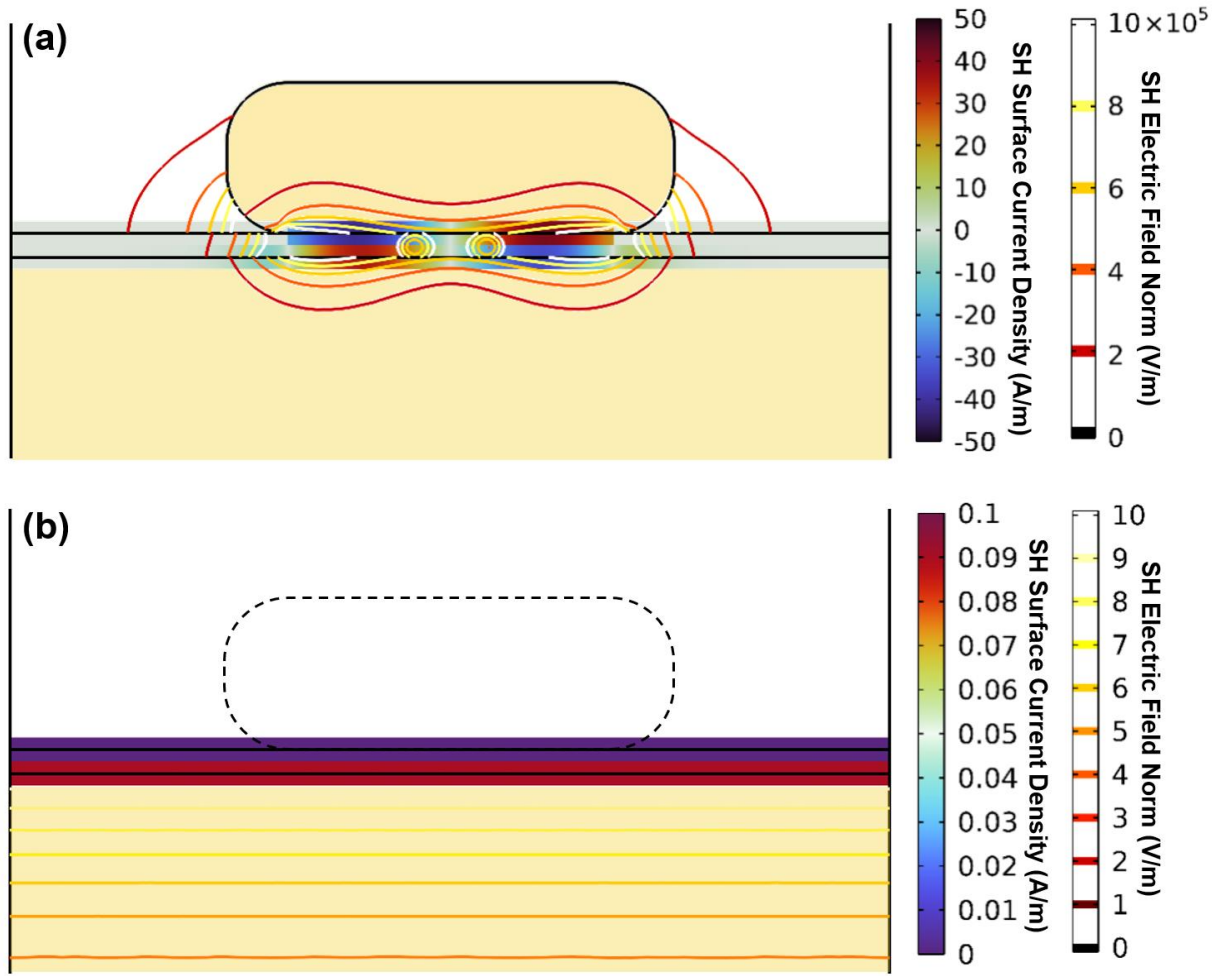

**Figure S5.** (a) Induced surface current density and generated electric field at SH wavelength when there is plasmonic nanocavity, computed by simulations. Note that the areas of strong surface current (in red and blue) are responsible for the generation of the boosted SH electric field which leads to enhanced SHG efficiency. (b) Similar results to (a) but without the nanostripe, i.e., no nanocavity formation. In this case, the induced surface current density and generated electric field at SH wavelength are orders of magnitude lower compared to the nanocavity results shown in (a), hence, leading to substantially lower SHG efficiency.